\documentclass[
    aps,
    prl,
    reprint,
    nofootinbib
]{revtex4-2}

\usepackage{graphicx}
\usepackage{amsmath}
\usepackage{mathtools}
\usepackage{xcolor}
\usepackage{bold-extra}
\usepackage[normalem]{ulem}
\usepackage{booktabs}
\usepackage{mleftright}
\usepackage{siunitx}
\usepackage[ISO]{diffcoeff}
\usepackage{placeins}
\usepackage{hyperref}
\usepackage[capitalise]{cleveref}

\makeatletter
\input{aas_macros.sty}
\let\jnl@style=\relax
\makeatother

\DeclareSIUnit{\Msun}{\text{\ensuremath{M_\odot}}}
\DeclareSIUnit{\pc}{\ensuremath{\mathrm{pc}}}

\definecolor{RedWine}{rgb}{0.743,0,0}
\definecolor{GrassGreen}{rgb}{0.125,0.75,0.125}
\definecolor{RoyalBlue}{rgb}{0.25,0.41,0.88}
\definecolor{DarkCyan}{rgb}{0,0.5,0.5}

\newcommand*{\vdis}{{\langle \mathbf{v}^2 \rangle}}
\newcommand*{\codename}{\textsc{KiSS-SIDM}}

\begin{document}

\title{Core Collapse Beyond the Fluid Approximation: The Late Evolution of Self-Interacting Dark Matter Halos}

\author{James Gurian}
\email{jgurian@perimeterinstitute.ca}

\author{Simon May}
\email{simon.may@pitp.ca}

\affiliation{Perimeter Institute for Theoretical Physics, Waterloo, Ontario, N2L 2Y5, Canada}


\begin{abstract}
We show that the gravothermal collapse of self-interacting dark matter (SIDM) halos can deviate from local thermodynamic equilibrium. As a consequence, the self-similar evolution predicted by the commonly adopted conducting fluid model can be altered or broken. Our results are obtained using a novel, efficient kinetic solver called {\codename} for tracing the gravothermal evolution based on the Direct Simulation Monte Carlo (DSMC) framework. In the long mean free path stage, the code is a viable alternative to the fluid model, yet requires no calibration parameters. Further, this method enables a fully kinetic treatment well into the late, short mean free path, stage of the collapse. We apply the method to a canonical case with isotropic, velocity independent scattering. We find that although a fluid treatment is appropriate deep in the short mean free path core, departures from local thermodynamic equilibrium develop in the intermediate mean free path region bounding the core, which modify the late-time evolution. {\codename} is publicly available at \url{https://gitlab.com/Socob/KiSS-SIDM}.
\end{abstract}

\maketitle


\section{Introduction}

Despite decades of research, the particle nature of dark matter (DM) remains unknown. In one class of models, (some of) the DM possesses an elastic scattering self-interaction, which is referred to as self-interacting dark matter (SIDM). This scattering transports thermal energy in SIDM halos. Initially, the energy transport thermalizes the DM in the center of the halo, leading to the formation of a core. Eventually, the outward energy transport from the hot core to the cool envelope of the halo leads to a runaway collapse, the ``gravothermal catastrophe''. This basic physical picture has been invoked to explain numerous astrophysical observations, including the diversity of galactic rotation curves and the formation of massive quasars at high redshift \cite{Balberg_2002, Kamada2017, Feng_2021, jiang2025formationlittlereddots}. However, obtaining a quantitative handle on the late evolutionary stages has proved challenging, and the endpoints of the collapse remain poorly understood. This is because the dynamics are described by the collisional Boltzmann equation, which is a formidable challenge for numerical methods due to the six-dimensional phase space.

Practically, computational approaches to this problem have bifurcated. When the mean free path (MFP) is long compared to the gravitational scale height (the LMFP regime), existing $N$-body codes have been modified to allow pairwise scattering between nearby particles \cite{2000ApJ...534L.143B, 2011MNRAS.415.1125K, rocha_cosmological_2013, Peter_2013, Tulin_2018}. This method becomes prohibitively expensive as the density increases and the MFP becomes short, because it requires identifying and calculating pairwise scattering probabilities between all nearby particles at each time step. Moreover, extensive recent research has emphasized the challenges of obtaining converged results and conserving the total energy of the system through the collapse \cite{palubski2024numericalchallengesmodelinggravothermal, Fischer_2024, mace2024convergencetestsselfinteractingdark, Meskhidze_2022}. 

Meanwhile, in the short mean free path (SMFP) limit, the behavior of the DM is treated by a moment approach to the Boltzmann equation, closed via an ideal gas equation of state {\cite{Moore_2000, Yoshida2000, Peebles_2000} and an effective thermal conductivity \cite{Balberg_2002, Essig_2019, Nishikawa_2020, Outmezguine:2022bhq, Gad-Nasr:2023gvf}}. The ``conducting fluid'' model has been adopted in the SMFP regime where the thermal conductivity can be derived from first principles \cite{Chapman70}. If the thermal conductivity is appropriately calibrated, the fluid approach can also produce good agreement with $N$-body simulations in the LMFP limit, at reduced computational cost. Recent work \cite{Yang_2023, mace2025calibratingsidmgravothermalcatastrophe} has focused on the details of this calibration procedure. However, the assumptions of the model have not been evaluated in the short and intermediate MFP regimes. Especially, the intermediate regime (where the MFP is comparable to the gravitational scale height) is problematic both for the fluid model and for $N$-body codes. In the former, an ad-hoc interpolation in the conductivity must be adopted, while in the latter the computational cost already becomes very high.


\section{Methods}

A standard technique in the computational fluid dynamics literature for approximate numerical solutions to the collisional Boltzmann equation is the Direct Simulation Monte Carlo (DSMC, \cite{bird1994molecular, Oran98, pareschi}) algorithm. As in astrophysical $N$-body codes, a finite number of discrete tracers are used to sample the continuous distribution function, and elastic collisions are simulated between pairs of tracer particles. However, in the DSMC approach simulation particles are grouped into {Eulerian} cells whose length is required to be smaller than the MFP. Scattering is allowed only between particles within the same cell. In the ``no time counter'' scheme which we adopt, at each time step and in each cell an upper bound on the number of collisions, $\Gamma_i$ {(defined in the End Matter)} is calculated. Then, for $\Gamma_i$ randomly chosen pairs of particles a rejection sampling is performed to determine if a collision occurs.

This approach enjoys several advantages compared to the standard neighbor search implementation in $N$-body codes. First, the cell to which each particle belongs can be stored and dynamically updated, obviating the neighbor search step. Second, it is guaranteed that no scattering will occur between particles separated by more than one {local} MFP. Finally, the number of trial collisions decreases in proportion to the time step, which greatly reduces the overhead of adopting a time step much smaller than the scattering timescale. This allows us to adopt a global (rather than per particle) time step, substantially improving the energy accuracy \cite{Fischer_2024}. {Still, the computational cost increases with the scattering rate and eventually becomes prohibitive. Various extensions to the method exist to control the computational cost in the frequent scattering limit \cite{fei2024navierstokes, pareschi, Weinberg_2014}.}

In our implementation, we assume spherical symmetry: all particles in the same spherical shell are allowed to scatter off of each other regardless of their angular coordinates. Therefore, the distribution function in each radial bin is extremely well resolved compared to a fully three-dimensional simulation with the same parameters and particle number. We likewise exploit this symmetry in calculating the gravitational acceleration. {This assumption is justified especially in the later stages of the collapse, as viscosity will efficiently remove angular momentum \cite{Feng_2021} and scattering relaxes the distribution to the lowest-energy (spherical) configuration.}

More details about the code, which we call {\codename} (``Kinetic Spherically Symmetric SIDM''), are provided in the End Matter. We have checked that {\codename} reproduces the SMFP Chapman–Enskog thermal conductivity to within \SI{2}{\percent}. The energy is conserved in the fiducial halo case discussed below to $|\Delta E/E| < \num{2e-4}$. {We have verified that the fiducial NFW halo simulated with collisionless CDM remains static to within statistical fluctuations for at least $80$ dynamical times ($t/t_0 \sim 200$), which is long compared to the core formation time here.} The method is computationally inexpensive in the short and intermediate MFP regimes: the core formation and early collapse results are converged with $\sim 10^5$ particles, an order of magnitude fewer than in \cite{mace2024convergencetestsselfinteractingdark}, and require roughly 30 minutes on a laptop.


\section{Results}

\begin{figure*}
    \centering
    \includegraphics[width=\linewidth]{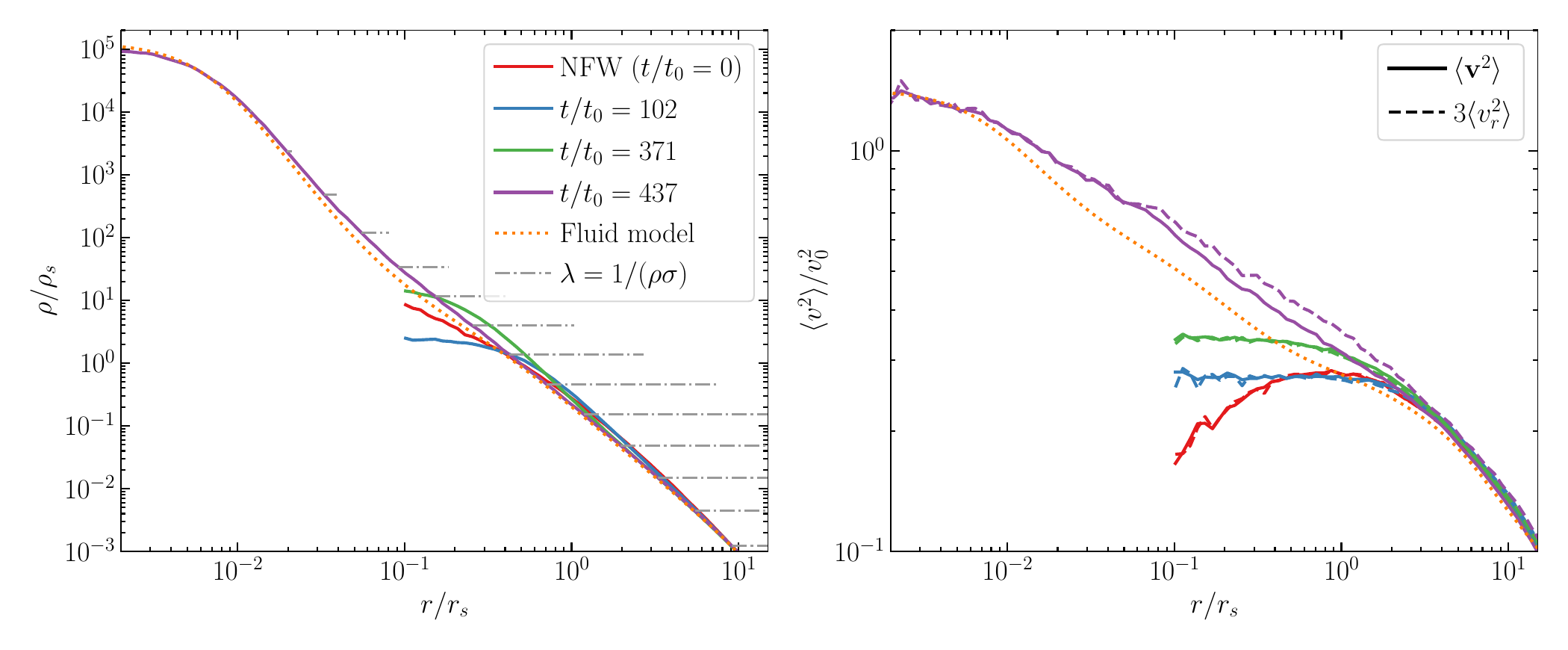}

    \caption{\textit{Left:} The density profiles in the simulation at different times (solid) and at a matched central density (at the final time) in the fluid model (dotted). The MFP at different densities is also plotted (length of the dashed lines). \textit{Right:} The velocity dispersion profiles for the same cases, 3D (solid) and triple the radial velocity dispersion (dashed). In the fluid model, the velocity dispersion is isotropic.}
    \label{fig:profiles}
\end{figure*}

We simulate an initially Navarro–Frenk–White (NFW) halo \cite{1997ApJ...490..493N} consisting of \num{2e6} particles. A constant cross section large enough to collapse the core of typical galaxy-sized halos within the age of the universe is likely excluded by observations \cite{adhikari2022astrophysical}. However, studies which consider the presence of baryons \cite{Feng_2021}, tidal stripping \cite{Nishikawa_2020}, or a velocity-dependent cross section \cite{Turner_2021} find runaway collapse by the present day. Here, we examine the canonical case of a constant cross section in an NFW halo in order to clearly illustrate the physics of the gravothermal collapse. Specifically, the density profile is:
\begin{equation}
    \rho(r) = \frac{\rho_{\mathrm{s}}}{\left(1+\frac{r}{r_{\mathrm{s}}}\right)^2\left(\frac{r}{r_{\mathrm{s}}}\right)},
\end{equation}
with $\rho_{\mathrm{s}}$ the scale density and $r_{\mathrm{s}}$ the scale radius. From the scale radius and density we define $M_0 = r_{\mathrm{s}}^3 \rho_{\mathrm{s}}$, $v_0 = \sqrt{G M_0/r_{\mathrm{s}}}$, $t_0^{-1} = a \sigma_m v_0 \rho_{\mathrm{s}}$ and $\sigma_0^{-1} = \rho_{\mathrm{s}} r_{\mathrm{s}}$, which we adopt as our units. 
Here, $G$ is the gravitational constant, $\sigma_m$ the cross section per unit mass, and $a = 4/\sqrt{\pi}$ enters the mean scattering time for hard sphere collisions.

We consider a case with $\sigma_m/\sigma_0 = 0.32$ (corresponding to, for example, a \SI{e9}{\Msun} halo with $\rho_{\mathrm{s}}=\SI{2.73e-2}{\Msun\per\pc\cubed}$, $r_{\mathrm{s}} = \SI{1.18e3}{\pc}$ and a cross section $\sigma_m = \SI{50}{\cm\squared\per\g}$ – the same as the low-concentration halo of \cite{palubski2024numericalchallengesmodelinggravothermal}), which we initialize using \textsc{SpherIC} \cite{Garrison_Kimmel_2013}. {We initialize the DSMC grid by 21 logarithmically spaced right-hand cell edges from $r/r_s\approx0.017$ (that is, $\SI{20}{\pc}$) to $r/r_s\approx1169$ (which is 10 times the initial radius of the outermost particle). At the final snapshot, the width of the smallest grid cell is $r/r_s \approx 1.66\times 10^{-5}$. The grid refinement procedure is detailed in the End Matter.} 

The resulting density and velocity dispersion profiles are shown in \cref{fig:profiles} at several times, along with the conducting fluid result with the central density matched to our final snapshot. \Cref{fig:profiles} shows that at the end of our simulation the MFP is very small compared to $\rho/|\nabla \rho|$, which indicates that the simulation has reached the SMFP regime. We have plotted both the 3D velocity dispersion $\vdis$ and thrice the radial velocity dispersion, $3 \langle v_r^2\rangle$. Here, we see that in the intermediate mean free path (IMFP) regime the radial velocity dispersion is modestly enhanced compared to the angular components and all components are enhanced compared to the conducting fluid model \cite{Gad-Nasr:2023gvf}. {At lower densities, the fluid and DSMC results agree well, and are not plotted to mitigate visual clutter.} {The development of radial velocity anisotropy is consistent with similar studies of globular clusters (which can undergo gravothermal collapse due to two-body gravitational scatterings) \cite{Giersz1994, takahashi1995fokkerplanck}}. {While this work was under review, \cite{fischer2025accuratelysimulatingcorecollapseselfinteracting} independently confirmed the anisotropy in the distribution function.}

\begin{figure*}
    \centering
    \includegraphics[width=\linewidth]{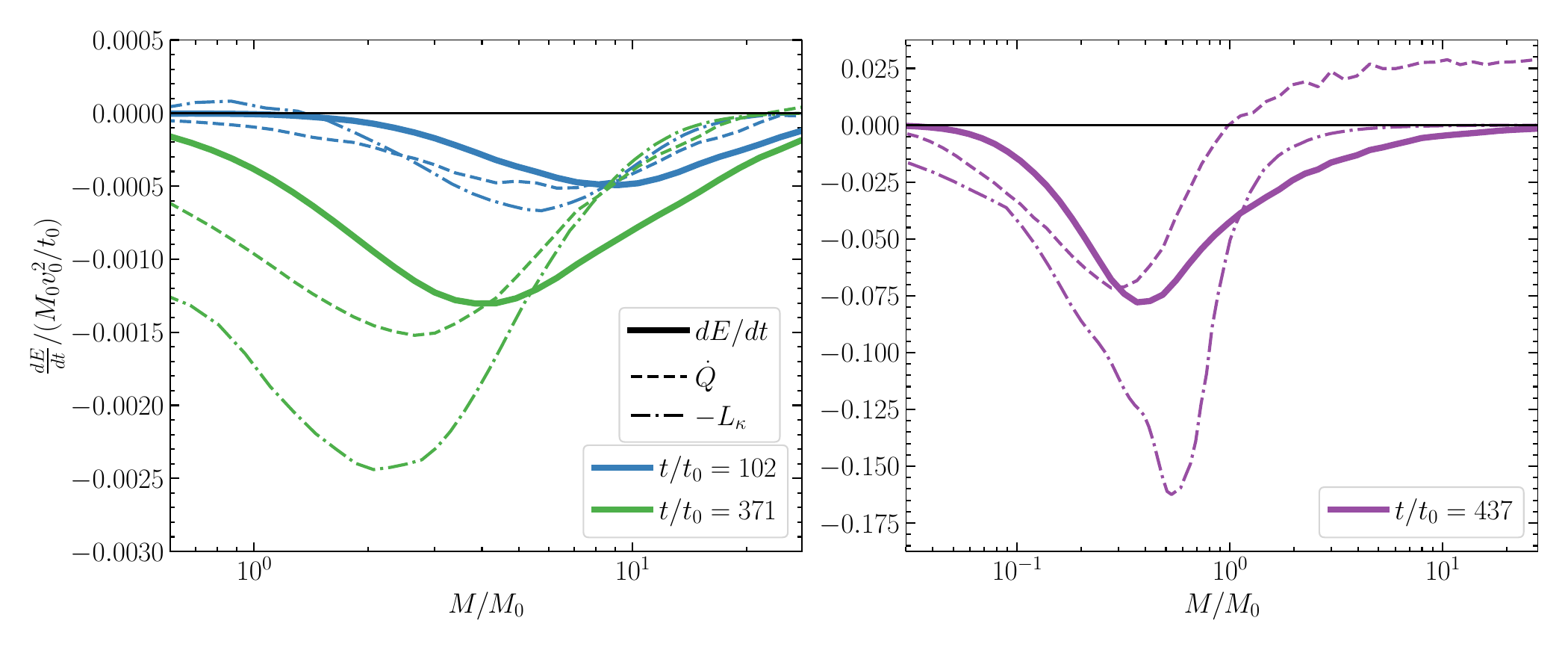}

    \caption{The {Lagrangian derivative of the energy (solid)}, the mass-integrated, temperature-weighted time derivative of the specific entropy (dashed), and the theoretical conductive luminosity (dash-dotted) at three points in the evolution, corresponding to \cref{fig:profiles}. The late evolution is shown on a separate panel due to the differing scales of interest.}
    \label{fig:Tds}
\end{figure*}

The fluid model closes the Boltzmann hierarchy by adopting an ideal gas equation of state and Fourier's law of thermal conductivity. This is similar to the treatment of proto-stellar evolution (e.\,g.~\cite{Stahler1980a, Stahler1980b, Hosokawa2009}). A similar model has also been applied to the gravothermal evolution of globular clusters, e.\,g.~\cite{Lynden-Bell80}.

Specifically, the second moment of the Boltzmann equation yields, for an ideal gas, the energy equation \cite{Balberg_2002}:
\begin{equation}
    - \left(\frac{\partial L}{\partial M}\right)_t = \frac{\vdis}{3} \left(\frac{\dl s_{\mathrm{I}}}{\dl t}\right)_M 
    \label{eq:energy}
\end{equation}
with $L$ the luminosity, $M$ the (enclosed) mass, and $s_{\mathrm{I}} = \ln(\langle \mathbf{v^2}\rangle^{3/2}/\rho)$ the specific entropy (up to an additive constant) of a monatomic ideal gas. The factor of three relates the 3D velocity dispersion $\langle\mathbf{v}^2\rangle$ to the temperature. In the presence of anisotropic (viscous) stress, this equation must be amended:
\begin{equation}
    -\left(\frac{\partial L}{\partial M}\right)_t + \pi_{ij}\frac{\partial u_i}{\partial x_j}= \frac{\vdis}{3} \left(\frac{\dl s_{\mathrm{I}}}{\dl t}\right)_M , 
    \label{eq:energydisip}
\end{equation}
where $\pi_{ij} = -\langle v_i v_j\rangle + \frac{1}{3}\vdis\delta_{ij}$ and $\mathbf{u} = \langle\mathbf{v}\rangle$ is the bulk velocity.

\begin{figure}
    \centering
    \includegraphics[width=\linewidth]{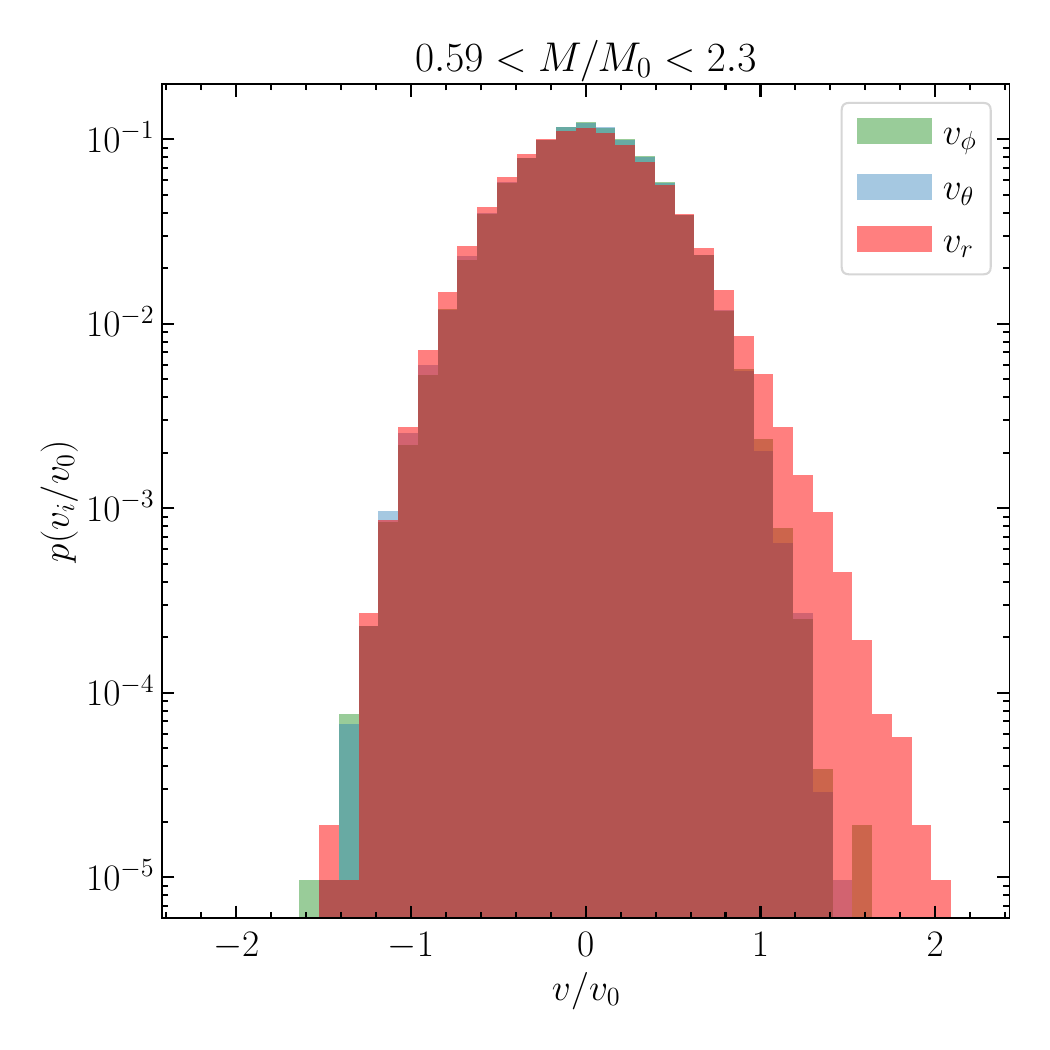}

    \caption{The velocity distribution function at a late time ($t/t_0 = 437$) in the IMFP region.}
    \label{fig:vhist}
\end{figure}

\enlargethispage{2ex}

Due to the anisotropy in the velocity distribution function  (\cref{fig:profiles}), we should expect energy transport by viscous dissipation in addition to conduction. Similar to \cite{Yang_2022}, we examine this hypothesis directly by comparing the conductive luminosity, the temperature-weighted entropy production rate. In the conducting fluid model, the following form is adopted for the luminosity:
\begin{equation}
    \frac{L_\kappa}{4 \pi r^2} = -\frac{1}{2} a \sqrt{\frac{\vdis}{3}} \left( \frac{12 \pi G}{C\rho \vdis\sigma_m} + \frac{a}{b} \sigma_m \right)^{-1}\frac{\partial \vdis}{\partial r},
    \label{eq:conduct}
\end{equation}
with $a = 4/\sqrt{\pi}$, $b=25\sqrt{\pi}/32$, and $C$ an unknown constant of order unity. The first term in parentheses is an approximate form of the LMFP thermal conductivity \cite{Lynden-Bell80, Balberg_2002}, while the second term can be exactly derived in the SMFP limit \cite{Chapman70}. Here, we adopt $C = 0.84$ (as in \cite{palubski2024numericalchallengesmodelinggravothermal}, which is similar to the case studied here). We interrogate the fluid model quantitatively by comparing the luminosity with the integral of the right-hand side of \cref{eq:energy},
\begin{equation}
    \dot{Q} \equiv \int \dl M \, \frac{\vdis}{3} \left(\frac{\dl s_{\mathrm{I}}}{\dl t}\right)_M,
\end{equation}

{Similar to \cite{Yang_2022}, we also calculate the Lagrangian derivative of the energy at mass coordinate $m$, which includes energy transfer through conductive luminosity, viscous dissipation, and mechanical ($P \dl V$) work:}
\begin{equation}
    \frac{\dl E}{\dl t} = \frac{\dl{}}{\dl t}\int_0^m \dl m' \left(-\frac{G M\, }{r} + \frac{\vdis}{2}\right),
    \label{eq:dedt}
\end{equation}
{with $M$ the enclosed mass at $m'$.} Accordingly, both sides of \cref{eq:energy} integrated over mass, are shown in \cref{fig:Tds} {along with the energy derivative}. In blue in the left panel, the entire halo is in the LMFP regime and {all three} expressions agree well, {indicating that the energy transfer is dominated by conduction of the form given by \cref{eq:energy}}. In green the core has entered the IMFP regime and shows worse agreement. In purple (right panel) the two sides of \cref{eq:energy} approximately agree in the SMFP core, {but the total energy loss is smaller in magnitude than these expressions because of compressional heating of the core by the rapidly cooling IMFP region.} In the IMFP regime the agreement is again poor, and we have an asymptotic value of $\dot{Q} > 0$, which indicates a breakdown of the conducting fluid model: equating this quantity with the luminosity would require heat flow against the temperature gradient. The deviation is naturally explained by viscous dissipation sourced in the IMFP regime

In \cref{fig:vhist} we show directly the velocity distribution in the IMFP regime. Here, the outward heat flux in the radial direction has not fully thermalized, and the distribution function is not isotropic. {This is because in this regime particles with a large, outward radial velocity can ``boil off'' into the low-density part of the halo, where they thermalize only after traveling a substantial distance.} Indeed, {local thermodynamic equilibrium (LTE), which we operationally define as the statement that at each spatial point the three-dimensional velocity distribution function is thermal (here, Maxwell–\allowbreak Boltzmann)}, is a fundamental assumption of the fluid model: \cref{eq:conduct} assumes a well-defined local temperature while \cref{eq:energy} omits the viscous dissipation associated with an anisotropic velocity distribution function. 

\begin{figure}
    \centering
    \includegraphics[width=\linewidth]{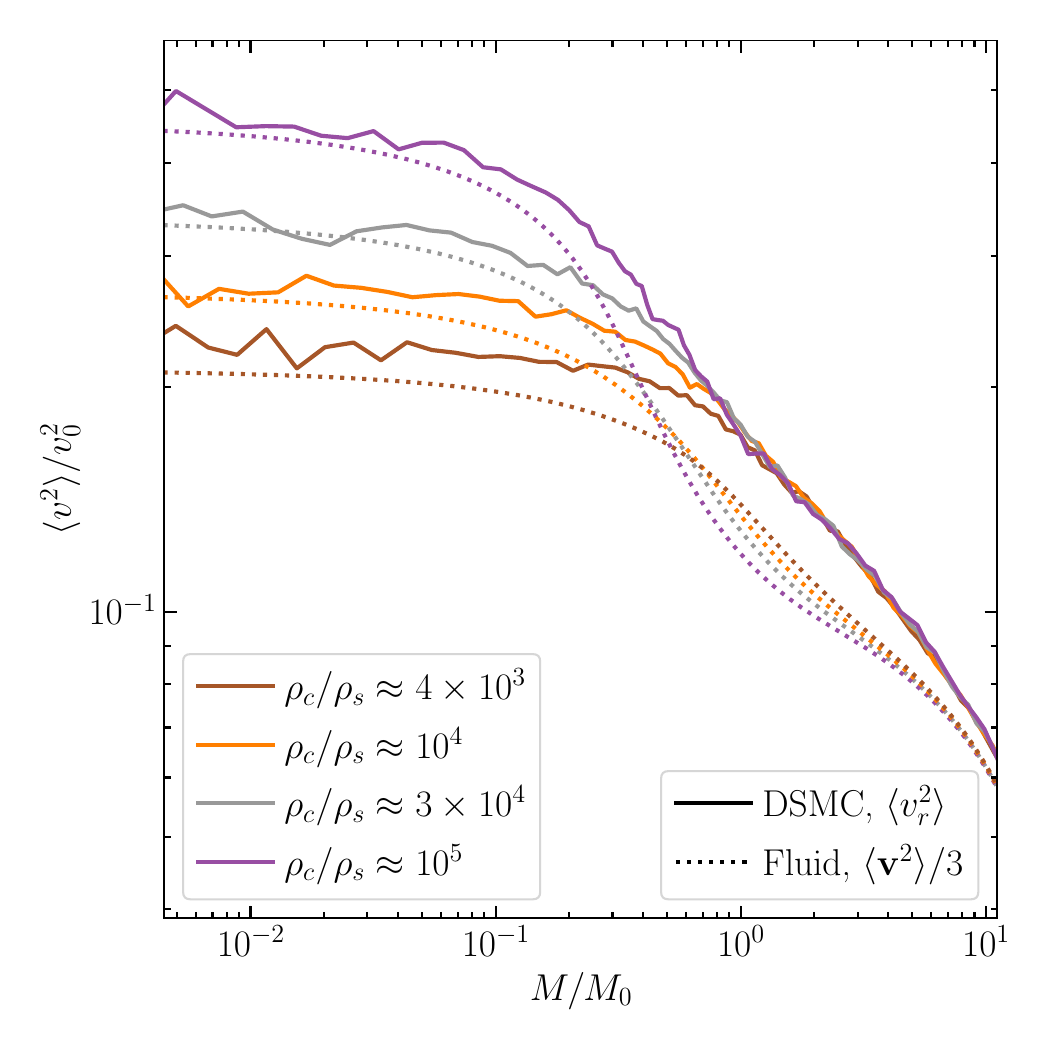}

    \caption{The DSMC radial velocity dispersion (solid) and fluid 1D velocity dispersions (dashed) at a sequence of central densities after the core enters the SMFP regime.}
    \label{fig:core_evo}
\end{figure}

Although LTE is established deep in the core, the structure of the core (i.\,e.~the slope of the density and velocity dispersion profiles near the core boundary) remains sensitive to the IMFP behavior. This is because the luminosity peaks in the IMFP regime, such that the evolutionary (Kelvin–Helmholtz) timescale hardly varies as a function of mass enclosed between the core and the IMFP region. In other words, the IMFP region is evolving on a similar timescale to the core. \Cref{fig:core_evo} shows these differences between the DSMC and fluid results at matched central densities. Although the central velocity dispersions are similar, the evolution outside of the core is qualitatively different between the two treatments. In the fluid model the core boundary becomes increasingly sharp over time as the velocity dispersion increases in the inner region and decreases in the outer region. Meanwhile, in the kinetic treatment the viscous heating leads to very slow evolution in the velocity dispersion outside the core.

\section{Discussion}

We have developed the new Direct Simulation Monte Carlo code {\codename} to study the gravothermal evolution of self-interacting dark matter halos. By exploiting the spherical symmetry in the idealized collapse problem and implementing a more efficient scattering algorithm, our code improves substantially on the speed and accuracy of previous $N$-body methods. This allows us to simulate the collapse well into the short mean free path regime using a fully kinetic method. 

In the IMFP region, we find qualitative disagreement with the fluid model due to the non-equilibrium state of the DM. In the canonical case studied here most of the evolutionary time is spent in the LMFP regime. Thus, the deviations from LTE result in only a negligible effect on the core collapse timescale and corresponding observational constraints. However, recent work has identified differences of $\sim \SI{30}{\percent}$ in the collapse time and minimal core density between different $N$-body SIDM implementations. In this context, the DSMC method will be an independent test and comparison point. Further, the non-equilibrium effects uncovered in this minimal SIDM scenario serve as a cautionary note for the study of more complex dark sectors containing both an elastic scattering self-interaction and other physics, such as dissipation (e.\,g.~\cite{Essig_2019}). 

On the other hand, there is long-standing interest in the possibility that core collapse in SIDM halos could explain the presence of supermassive black holes in galactic centers \cite{Balberg_2002, Pollack2015, Feng_2021, jiang2025formationlittlereddots, Gad-Nasr:2023gvf}. There, the key parameter is the core mass at the time when relativistic instability induces dynamical collapse, which these studies have investigated by extrapolating fluid model results to very high central densities. Our results confirm that deep in the SMFP regime LTE is established and the conducting fluid model is appropriate. Nonetheless, the non-equilibrium IMFP region remains relevant deep into the collapse.

Ref.\ \cite{Balberg_2002} argues that the mass enclosed within the IMFP region sets the mass of the final black hole, as interior to this scale the DM will respond hydrodynamically (i.\,e.~collapse) due to a loss of pressure support in the core, while in the LMFP region a new (nearly) collisionless equilibrium will be established. In the fluid model, the density and velocity dispersion profiles in the IMFP region continue to evolve even when the core is deep into the SMFP evolution. {In particular, in the fluid treatment the SMFP core eventually develops a sharp ``wall'' with hardly any mass in the IMFP regime, which is beginning to be visible in \cref{fig:core_evo}. The evolution of the core mass is then a robust, well defined quantity which has been described in terms of simple power-law scalings \cite{Balberg_2002, Gad-Nasr:2023gvf, Outmezguine:2022bhq}. In the DSMC results, the behavior is qualitatively different, such that the appropriate definition of the core is unclear. We discuss this issue further in the End Matter.}

The difference between the fluid model and our simulation result is still increasing at the end of our simulation, while the central density and velocity dispersion remain many orders of magnitude below the relativistic instability threshold. Thus, it is dangerous to study the core structure at late times using the fluid model. {If the trend of the IMFP region being continually ``refilled'' in our results holds all the way to the relativistic instability, the final black hole mass would be much larger than previous estimates.}

Similarly, in the fluid model the evolution is ``universal'' in that $\sigma_m / \sigma_0$ is the only parameter controlling the dynamics in initially NFW halos, but it is possible that the viscous heating identified here introduces a parametrically different time dependence and hence a broader range of collapse behaviors. In the future, it will be particularly interesting to apply the DSMC method to anisotropic, velocity-dependent cross sections. {Generalizing the DSMC algorithm to this case is technically straightforward, although some care is required in the definition of $\Gamma_i$ to avoid an excessively large number of trial collisions.} Because a large cross section is not observationally excluded at low velocities, the halo may enter the IMFP regime at earlier times and lower densities. Moreover, both the high-temperature cutoff in the cross section and a differential cross section favoring forward/\allowbreak backward scattering may substantially impede thermalization of the distribution function at high densities. 

{Finally, we comment that despite its name, {\codename} was developed with the generalization to three dimensions in mind. While appropriately updating the cell-assignment logic and gravity solver is non-trivial, it is reasonable within the extant framework. However, the fundamental problem that time step and resolution demands become intractable in regions of extreme density contrast will persist. Correct implementation of core collapse in cosmological simulations will require developing subgrid prescriptions motivated by highly accurate simulations in idealized settings. {\codename} will play an integral role in this program.}

\begin{acknowledgments}
\paragraph*{Acknowledgments.}
We thank Neal Dalal, Moritz Fischer, Laura Sagunski, and Kimberly Boddy for useful discussions. We thank Michael Ryan for sharing a modified version of the \textsc{SpherIC} initial conditions generator. Research at the Perimeter Institute is supported in part by the Government of Canada through the Department of Innovation, Science and Economic Development Canada and by the Province of Ontario through the Ministry of Colleges and Universities.
\end{acknowledgments}

\bibliographystyle{aasjournal}
\bibliography{sidm.bib}

\clearpage

\onecolumngrid
\section*{End Matter}
\twocolumngrid

\appendix


\paragraph{\hspace*{-\parindent}\bfseries Details of the Direct Simulation Monte Carlo Code {\codename}:} In our DSMC algorithm, {\codename}, space is discretized into an {adaptively split Eulerian} grid whose cells are smaller than the mean free path, and scattering between particles can only occur within a single cell. For the no time counter scheme, the following upper bound on the number of collisions in each cell is calculated at each time step $\Delta t$:
\begin{equation}
    \Gamma_i = N_i \rho_i\sigma_{m,\mathrm{max},i} v_{\mathrm{max},i} \Delta t,
\end{equation}
where $N_i$ is the number of tracers in the cell $i$, $\rho_i$ is the mass density in this cell, $\sigma_{m, \mathrm{max},i}$ is an upper bound on on the cross section (per unit mass),  $v_{\mathrm{max},i}$ is an upper bound on the relative velocity between particles, and $\Delta t$ the time step, {discussed below}. 

At each time step, and in each cell $\Gamma_i$ pairs (trial pairs) are sampled. Each trial pair of particles collides if:
\begin{equation}
    \frac{\sigma_m(v_{\mathrm{rel}}) v_{\mathrm{rel}}}{\sigma_{m,\mathrm{max},i} v_{\mathrm{max},i}} > q,
\end{equation}
where $\sigma_m$ is the cross section per unit mass and $q$ is randomly drawn from a uniform distribution on $[0,1)$. Multiple scatterings of the same particle pairs are not explicitly forbidden. This issue of repeated scattering becomes negligible when the number of particles per cell is large: \cite{AKHLAGHI2018} mention a rule of thumb that when this number exceeds 10 repeated scatterings do not require any special handling. This condition is satisfied in our simulations. {Collisions do not exactly conserve the pairwise angular momentum, but the expectation of the angular momentum in each cell is conserved, and the variance is controlled by the cell size \cite{Nanbu88}. Our spherically symmetric treatment is not generally appropriate for simulating cases with non-zero total angular momentum.}

{The time step is shared by all particles and is set adaptively as }
 \begin{equation}
    \Delta t = \min(\Delta t_{\mathrm{coll}}, \Delta t_{\mathrm{grav}}),
\end{equation}
where the collision time step is
\begin{gather}
    \Delta t_{\mathrm{coll}} = \min_i\mleft(\frac{\lambda_{\mathrm{MFP},i}}{v_{\mathrm{max},i}}\mright)
\shortintertext{with}
    \label{eq:mfp}
    \lambda_{\mathrm{MFP},i} = \frac{1}{\rho_i \sigma_{m,\mathrm{max},i}}
\end{gather}
is an estimate of the mean free path within cell $i$. {Meanwhile, the gravitational time step is determined via}
\begin{gather}
    \Delta t_{\mathrm{grav}} = \epsilon \min_i \mleft(t_{\mathrm{ff},i}\mright),
\intertext{where we choose $\epsilon = \num{0.02}$ and}
    t_{\mathrm{ff},i} = \frac{1}{\sqrt{G \bar{\rho}_i}}
\end{gather}
is the free-fall time in grid cell $i$, with the average density $\bar{\rho}_i$ within the sphere that reaches the center of the cell.

We assume spherical symmetry by recording only the radial position coordinate $r$ and the velocity components $\mathbf{v} = (v_r, v_\phi, v_\theta)$. At each time step the 3D (Cartesian) coordinates of each particle are first updated as 
\begin{equation}
    \mathbf{x'} = r \mathbf{\hat r} + \mathbf{v}\Delta t.
\end{equation}
Then the new angular basis is formed from $\mathbf{\hat r'} = \mathbf{x'}/|\mathbf{x'}|$ and a pair of orthonormal tangential vectors. The velocity components are rotated into this new $\mathbf{\hat r'}$, $\mathbf{\hat \phi'}$, $\mathbf{\hat \theta'}$ basis and stored. Finally, the new radial coordinate is calculated as $r' =|\mathbf{x'}|$ and stored.

All particles within one spherical shell (grid cell) can scatter. The gravitational force is simplified by spherical symmetry. The gravitational acceleration on each particle is
\begin{equation}
    a_{\mathrm{g}} = \frac{G M_{\mathrm{enc}}}{r^2},
\end{equation}
where $M_{\mathrm{enc}}$ is the mass interior to the particle. Within each cell, we smooth the mass distribution (and reduce the computational cost) by approximating the cumulative mass function as locally log-linear: $M(r) = M_i (r/r_i)^\alpha$, where $M_i$ is the mass enclosed at the lower boundary $r_i$ of the cell, and $\alpha$ is chosen to ensure continuity of the cumulative mass enclosed between cells. Note that the grid used for the collision and gravity steps does not necessarily need to be the same; however, we have chosen to identify the two for simplicity in our code.

As the simulation proceeds, the mean free path within each cell, as well as the required spatial resolution for the gravitational accelerations evolves. In order to ensure that the grid is fine enough at all times, we split grid cells into two once the {local} mean free path (\cref{eq:mfp}) becomes smaller than the grid cell. Similarly, we split cells if the fraction $f_{\mathrm{J}}$ of the Jeans length $\lambda_{\mathrm{J},i}$ within the cell,
\begin{equation}
    \lambda_{\mathrm{J},i}
    = \sqrt{\frac{\vdis_i}{G \rho_i}},
\end{equation}
becomes smaller than the cell size (choosing $f_{\mathrm{J}} = 1/20$). However, cells are only split if they contain a minimum number of particles, $n_{\mathrm{min}} = 32$.

For the time integration, we divide each time step as 
\begin{equation}
    K(\Delta t/2) \, S(\Delta t/2) \, D(\Delta t) \,S(\Delta t/2) \, K(\Delta t/2),
\end{equation}
where $K$ is the gravitational kick, $D$ the position update (``drift''), and $S$ the scattering. This formulation is not exactly time reversible due to the stochasticity of the scattering and the non-reversibility of the time step, yet produces good long-term accuracy. The code makes use of shared-memory parallelization (threading).

\begin{figure}
    \centering
    \vspace*{-1cm}
    \includegraphics[width=\linewidth]{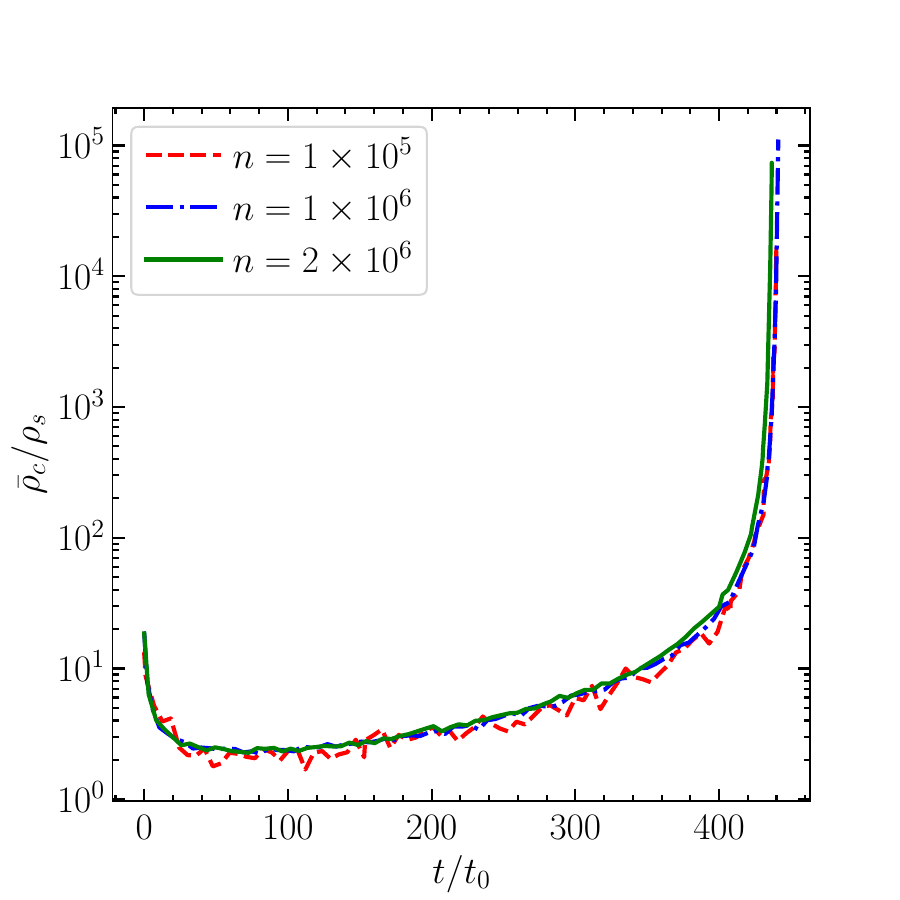}

    \caption{The time evolution of the mean density in a central region with differing numbers of particles.}
    \label{fig:time_evol}
\end{figure}

\paragraph{\hspace*{-\parindent}\bfseries Verification:}
{We have performed various tests of the code in idealized test problems, including measuring the thermal conductivity in the Chapman–Enskog limit, and 1D shock tests at varied Knudsen numbers. The time step and grid size conditions described above were determined and validated by these tests. Further, we have checked that an isothermal, cored density profile remains static for at least tens of dynamical times in the collisional and collisionless cases.} {We have rerun the fiducial simulation ($2\times 10^6$ particles) with $10^5$ and $10^6$ particles. \Cref{fig:time_evol} shows the time evolution of the mean density $\bar\rho_c$ calculated at $M/M_0 = 0.0334$, which for the lowest-resolution simulation corresponds to the innermost 100 particles. The core collapse times agree to within a few percent. We have also verified that the density and velocity dispersion profiles agree, although the results in the lowest resolution simulation are noisy. } {We have additionally verified that the results are insensitive to the details of the grid refinement procedure, as long as the local mean free path and Jeans length are resolved while ensuring a minimum number of particles per cell.}

\paragraph{\hspace*{-\parindent}\bfseries Core Mass Scaling:}
{Ref.~\cite{Balberg_2002a} define a Knudsen number by the ratio of the mean free path to the gravitational scale height:
\begin{equation}
    \mathrm{Kn} = \sqrt{\frac{\vdis}{12 \pi G \rho}} \frac{1}{\rho\sigma_m}
    ,
\end{equation}
and then define the core mass by the scale where $\mathrm{Kn} = 1$ (note that this definition differs from \cite{Gad-Nasr:2023gvf}), which they use to calculate $\dl\log(M_{\mathrm{Kn}=1}) / \dl\log(\vdis) \approx -0.85$. The approach is reasonable because at  late times in the fluid treatment the core boundary is very sharp, so that the core mass and its scaling do not depend sensitively on the exact value of $\mathrm{Kn}$ used to define the core. At the conclusion of our simulation, the fluid results have not fully reached this phase of the evolution, while the DSMC results show a much less dramatic ``hollowing out'' in the density profile and no evidence of this behavior in the velocity dispersion profile. Therefore, it is not clear how the core should be defined, or whether its mass scales as a power law all the way to the relativistic instability. Still, we calculate the local power law best fits to the mass enclosed within $\mathrm{Kn} = 1$ and $\mathrm{Kn} = 5$, over the range $10^4 < \rho/\rho_{\mathrm{s}} < 10^5$ in \cref{tab:fits}. We find order one disagreement in this exponential scaling. The fluid scaling is much shallower than $-0.85$ because $\mathrm{Kn} = 1$ falls at $M/M_0 \sim 0.5$, which at this epoch is near the fixed point of the profile (\cref{fig:core_evo}).}

\begin{table}
    \centering
    \caption{Fits to the core mass scaling.}
    \smallskip

    \begin{tabular}{l S[table-format=+1.2] S[table-format=+1.2]}
        \toprule
         & {$\displaystyle \frac{\dl \log(M_{\mathrm{Kn}=1})}{\dl \log(\vdis)}$} & {$\displaystyle \frac{\dl\log(M_{\mathrm{Kn}=5})}{\dl\log(\vdis)}$} \\
         \midrule
         Fluid & -0.27 & -0.37 \\
         DSMC  & -0.21 & -0.21 \\
         \bottomrule
    \end{tabular}
    \label{tab:fits}
\end{table}

\paragraph{\hspace*{-\parindent}\bfseries Technical Comments:}
The curves in \cref{fig:profiles,fig:Tds,fig:core_evo} are averages over closely spaced snapshots to control the noise due to limited particle number. Our use of the integrated quantities $L$ and $\dot{Q}$ avoids noisy numerical derivatives required to evaluate both sides of \cref{eq:energy} directly in the simulation data. Still, \cref{eq:conduct} requires a spatial derivative of the velocity dispersion, which leads to artifacts in the core region, where the gradient is small. {Ref.~\cite{Yang_2022} equated the total derivative of the energy with the conductive luminosity, which neglects the mechanical work exchanged between mass elements. In principle, the conductive luminosity can be directly evaluated from simulation data, but as the third moment of the distribution function it is noisy in practice. } In \cref{fig:vhist}, we have not subtracted the bulk velocity, in order to emphasize that the radial distribution function has a mode near zero and differs from the angular components in the tails.

{Ref.~\cite{fischer2025accuratelysimulatingcorecollapseselfinteracting} found that the core is well fit by a King profile, 
\begin{equation}
    \rho(r) = \rho_0 \left[1 + \left(\frac{r}{r_{\mathrm{c}}}\right)^2\right]^{-3/2}.   
\end{equation}
We confirm that this profile is a good fit to the core in the LMFP and IMFP regimes, but find poor agreement in the SMFP regime. There the true density profile is somewhat steeper in the innermost region and much shallower in the outer region than the best-fit King profile.}

\clearpage

\end{document}